\documentclass[%
 reprint,superscriptaddress,groupedaddress
 amsmath,amssymb,
 aps]{revtex4-2}

\usepackage{graphicx}% Include figure files
\usepackage{dcolumn}% Align table columns on decimal point
\usepackage{bm}% bold math
\usepackage{siunitx}
\usepackage{braket}
\usepackage{amsmath,amssymb,amsfonts}
\usepackage{hyperref}
\hypersetup{colorlinks=true,allcolors=blue}
\usepackage[table]{xcolor} % For row coloring
\usepackage{booktabs} % For better horizontal lines
\definecolor{lightgray}{gray}{0.9}
\usepackage{todonotes}

\newcommand{\fmix}[0]{f_{\mathrm{mix}}}
\newcommand{\TUdo}{Condensed Matter Theory, Department of Physics, TU Dortmund, 44221 Dortmund, Germany}

\begin{document}

\preprint{arxiv}

\title{Tunable multi-photon correlations from a coherently driven quantum dot}

\author{Thomas K. Bracht}
\email{thomas.bracht@tu-dortmund.de}
    \affiliation{\TUdo}    
    
\author{Rachel N. Clark}
    \affiliation{School of Engineering, Cardiff University, Queen's Buildings, The Parade, Cardiff, CF24 3AA, UK}
    \affiliation{Translational Research Hub, Maindy Road, Cardiff, CF24 4HQ, UK}

\author{Petros Androvitsaneas}
    \affiliation{School of Engineering, Cardiff University, Queen's Buildings, The Parade, Cardiff, CF24 3AA, UK}
    \affiliation{Translational Research Hub, Maindy Road, Cardiff, CF24 4HQ, UK}

\author{Matthew Jordan}
    \affiliation{School of Engineering, Cardiff University, Queen's Buildings, The Parade, Cardiff, CF24 3AA, UK}
    \affiliation{Translational Research Hub, Maindy Road, Cardiff, CF24 4HQ, UK}

\author{Samuel G. Bishop}
    \affiliation{School of Engineering, Cardiff University, Queen's Buildings, The Parade, Cardiff, CF24 3AA, UK}
    \affiliation{Translational Research Hub, Maindy Road, Cardiff, CF24 4HQ, UK}

\author{Harry E. Dyte}
    \affiliation{School of Engineering, Cardiff University, Queen's Buildings, The Parade, Cardiff, CF24 3AA, UK}
    \affiliation{Translational Research Hub, Maindy Road, Cardiff, CF24 4HQ, UK}

\author{Moritz Cygorek}
    \affiliation{\TUdo}
    
\author{Ian A. Farrer}
    \affiliation{Department of Electronic and Electrical Engineering, University of Sheffield, Mappin Street, S1 3JD, Sheffield, UK.}

\author{Doris E. Reiter}%
    \affiliation{\TUdo}
\author{Anthony J. Bennett}
    \email{BennettA19@cardiff.ac.uk}
    \affiliation{School of Engineering, Cardiff University, Queen's Buildings, The Parade, Cardiff, CF24 3AA, UK}
    \affiliation{Translational Research Hub, Maindy Road, Cardiff, CF24 4HQ, UK}

\date{\today}

\begin{abstract}
Mixing the fields generated by different light sources has emerged as a powerful approach for engineering non-Gaussian quantum states. Understanding and controlling the resulting photon statistics is useful for emerging quantum technologies that are underpinned by interference. In this work, we investigate intensity correlation functions arising from the interference of resonance fluorescence from a quantum emitter with a coherent laser field. We show that the observed bunching behavior results from a subtle interplay between quantum interference and the normalization of the correlation functions. We show that by adjusting the mixing ratio and phase one can achieve full tunability of the second-order correlation, ranging from anti-bunching to bunching. We further extend our analysis to third-order correlation functions, both experimentally and theoretically, to provide new insights into the interpretation of higher-order correlations and offer practical tools for shaping quantum optical fields.
\end{abstract}

\maketitle

%%%%%%%%%%%%%%%%%%%% Introduction %%%%%%%%%%%%%%%%%%%% 
\section{Introduction}
Measuring the quantum state of light is a fundamental requirement to reveal its quantum nature. From quantum optics textbooks we learn that only by using higher-order intensity correlation functions can we probe the photon statistics of light. In the case of light from a laser, the second order correlation is consistent with light consisting of photons with a Poissonian number distribution, i.e., a coherent state, with a constant $g^{(2)}(t)=1$.  However, for continuously-driven quantized transitions, the canonical metric to assess if the light is unambiguously ``quantum” \cite{gerry2023introductory} is when the second order correlation function $g^{(2)} (0)$ is close to zero. These concepts can be probed experimentally using for example single atoms \cite{weber2006analysis,vasilev2010single}, molecules \cite{brunel1999triggered,lounis2000single} or defect centers in diamonds \cite{aharonovich2009enhanced,schroeder2011fiber,orphal2023optically}. Here, we use single semiconductor quantum dots (QDs), positioned in monolithic cavities \cite{reithmaier2004strong,reitzenstein2007alas,gazzano2013bright,senellart2017high}. Such systems have high oscillator strength, near-unity internal quantum efficiency and wavelengths compatible with efficient single photon detectors, which allows for convenient multi-photon coincidence rates \cite{heindel2023quantum,rickert2025high}. With use of cavities and resonant driving, these dots can emit highly indistinguishable photons \cite{senellart2017high, androvitsaneas2023direct, tomm2021bright}. These systems also display the rich physics of the solid state environment which can enable novel quantum phenomena not yet seen in atom-cavity systems, such as phonon-assisted processes \cite{Hennessy2006,reiter2019distinctive} and single-emitter spontaneous parametric down conversion \cite{Liu2025}.

Of particular interest is the case of the continuous-wave (CW) resonantly driven two-level system, which can be used as a testbed to explore the physics of antibunching, Rabi oscillations and squeezing \cite{kimble1977photon,Flagg2009, Muller2007}. At Rabi frequencies far below the inverse natural linewidth, variously referred to as the Heitler regime or the coherent-scattering regime, the light generated has been repeatedly shown to display a component with the same spectrum of the laser and yet be simultaneously anti-bunched on a timescale determined by the transition \cite{matthiesen2012subnatural, Nguyen2011, Bennett2015}. These apparently paradoxical observations are now accepted to be the result of interference between the scattered (coherent) laser and the incoherent emission from the transition \cite{Brash2020, Hanschke2020}. In sum, the resonance fluorescence retains partial phase coherence with the driving laser, and they can be made to interfere to generate a wide range of autocorrelation values \cite{Tomm2024,steindl2021artificial,kim2024unlocking}. Quantum dot sources can also be operated under pulsed excitation for deterministic single photon generation \cite{michler2000aquantum,Somaschi2016,tomm2021bright}.
Besides the fundamental interest, mixing a deterministic source of single photons with a coherent state in a homodyne measurement has practical applications towards a universal scheme for continuous-variable photonic quantum computing, by introducing the crucial non-Gaussian resource via, for example, photon addition or subtraction \cite{lam2025optimizing,lloyd1999quantum,eaton2019non}. 

In this paper, we study $g^{(n)}(t)$ correlation functions both in theory and experiment from a neutral exciton in a single quantum dot, providing detailed guidance on how to understand and control the obtained signals. Using an experimental system that can correlate light from orthogonal and parallel polarizations we probe the simultaneous observations of bunching and antibunching in multi-photon coincidence events. After introducing the underlying theory, we scrutinize the different components leading to specific behavior of the $g^{(2)} (t_{12})$ and then extend our findings to measure and understand $g^{(3)} (t_{12}, t_{23})$. 

\section{Setup}
\subsection{Experiment setup}

\begin{figure}
	\centering
	\includegraphics[width=\linewidth]{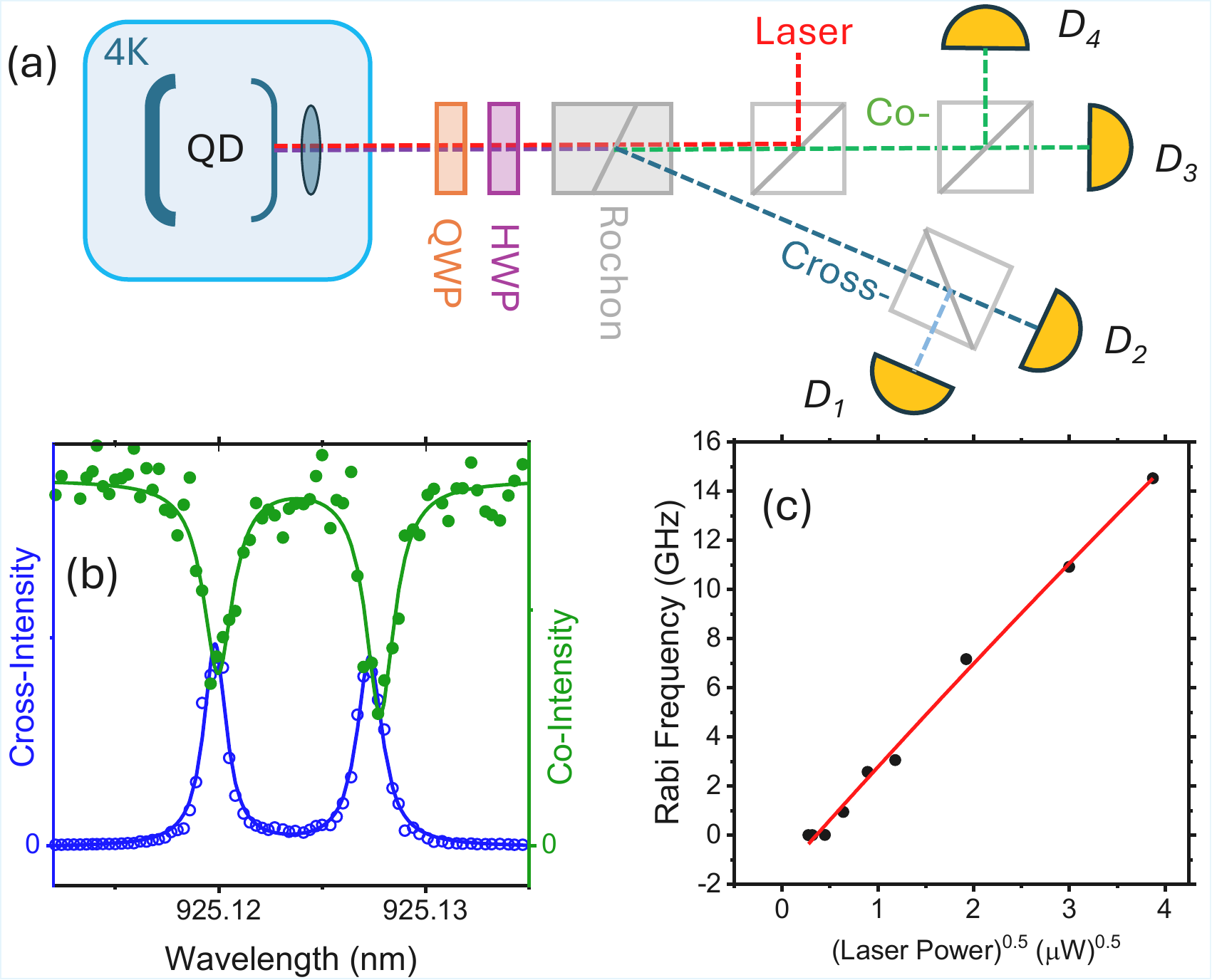}
	\caption{Experimental system to study multi-photon correlations from a resonantly driven quantum dot. (a) Schematic of the experimental system.  The tunable resonant laser is directed on to the unbalanced cavity containing the single quantum dot. Quarter and half- waveplates (QWP and HWP) are used to minimize laser scatter into the cross-polarized detectors whilst maximizing the visibility of the spectral features in the co-polarized detectors. (b) typical spectrum resulting from scanning the laser over the neutral exciton. (c) Rabi frequency varies linearly with the square-root of the incident laser power.} 
    \label{fig1}
\end{figure}

The experimental set-up used is shown schematically in Figure \ref{fig1}(a). Linearly polarized light from a narrow-band CW laser is focused onto the InAs QD embedded in an asymmetric semiconductor pillar microcavity with a quality factor of 430 $\pm$ 11 (more details can be found in \cite{androvitsaneas2023direct}). The linear polarization of the laser is set to equally couple to the two transitions of the neutral exciton, which are aligned to the primary axes of the GaAs crystal at approximately \SI{925.12}{\nano \meter}. Light collected from the sample is separated into two orthogonal polarizations by means of a Rochon prism before being directed to four superconducting single photon detectors, in the configuration shown in Figure \ref{fig1}(a). Sweeping the laser across the transitions results in a pair of Lorentzian peaks in the spectrum recorded by the cross-polarized detectors ($D_1$ and $D_2$), shown in Figure \ref{fig1}(b). Simultaneous detection of the light which is co-polarized with the laser is recorded by detectors $D_3$ and $D_4$, which show dips in the intensity corresponding to the wavelengths of the excitonic transitions. By adjusting the waveplates shown in Figure \ref{fig1}(a), it is possible to maximize the extinction of the laser on the cross-polarized channel and maximize the visibility of the features in the co-polarized spectrum. Autocorrelation histograms measured from the cross-polarized channels allow us to fit the Rabi frequency as a linear function of the square-root of incident laser power in Figure \ref{fig1}(c).  

In subsequent figures, the detectors have been used to record photo-detections on all four detectors simultaneously, when the laser is resonant with the lower wavelength transition. Offline post-processing with custom software can be used to generate all combinations of correlation histogram, such as $g^{(2)}(t_{ij})$ as a function of the times between detector $i$ and $j$. In addition, higher-order correlations $g^{(3)}(t_{ij},t_{ik})$ or $g^{(4)} (t_{ij},t_{ik}, t_{il})$ can be calculated between detectors $i, j, k$ and $l$. We ensure all detectors are within their linear-response regime so the correlations accurately sample the photon statistics of the system \cite{Clark2024}.
\subsection{Theoretical description}

In general, the normalized $n$th order coherence can be calculated using 
\begin{equation}
	g^{(n)}(t,\tau_{12},...) = \frac{G^{(n)}(t,\tau_{12},...)}{\braket{a^{\dagger}(t) a(t)}^n}
\end{equation}
where $G^{n}$ is the non-normalized $n$th order correlation function, which is normalized by the $n$th power of the detected intensity $\braket{a^\dagger a}$ at time $t$. Focusing on CW excitation, we assume that the system of QD and laser is in a steady state at time $t$. % , thereby, reducing the number of delays between the pulses to be taken into account by one.
For experimentally acquired data, the normalization is usually done with respect to the uncorrelated steady state value. The two-time correlation function reads explicitly
\begin{equation}
	G^{(2)}(\tau) = \braket{a^{\dagger}(t)a^{\dagger}(t+\tau)a(t+\tau)a(t)}\,. \label{eq:g2}
\end{equation}

It describes the amplitude of events given by the detection of a photon at time $t$ and a second photon at $t+\tau$.  Negative delays $\tau<0$ can be recreated in theory by using the appropriate time ordering for the operators or by symmetry considerations. Experimentally, negative delays are recorded by time-tagging each detection event and correlating the events afterwards.

In practice, calculation of the correlation functions means the system is evolved until it reaches a steady state at point $t$, where the operators with argument $t$ are applied and the quantum regression theorem (QRT) is used. Afterwards, the expectation value of the operators with $t+\tau$ yields $G^{(2)}(\tau)$. Higher-order correlation functions are calculated by repeated application of the QRT. Since we do not consider environmental effects such as the coupling to LA phonons, the QRT is exact \cite{cosacchi2021accuracy}. 

Analogously, the three-time correlation function reads
\begin{align}
\begin{split}
    G^{(3)}(t,\tau_{12},\tau_{13}) = \langle&a^{\dagger}(t)a^{\dagger}(t+\tau_{12})a^{\dagger}(t+\tau_{13})\times\\
    &a(t+\tau_{13})a(t+\tau_{12})a(t)\rangle,
\end{split}
\end{align}

describing the detection of a photon at time $t$, a second photon at $t+\tau_{12}$ and a third photon at $t+\tau_{12}$. This implies that in theory $\tau_{13}\ge\tau_{12}$, otherwise the results are not physical \cite{delvalle2012theory,bracht2024theory}. 

In the experiment, we discriminate between different cases. Without any laser mixing, the photon modes $a/a^{\dagger}$ become the quantum emitter (QE) transition operators $\sigma^{\dagger}$ and $\sigma$ \cite{cosacchi2021accuracy}. These operators describe the excitation of the system from the ground to the excited state and vice versa. This gives rise to the $G_{\text{cross},\text{cross}}$ signal, that is typically calculated for a QE without laser mixing.

To describe the mixing of QE signal with the coherent state of the laser, we extend QE transition operators to \cite{zubi2020tuning}
\begin{equation}
	s = \sigma + \beta = \ket{g}\bra{x} + \fmix \braket{\sigma} e^{i\phi}. \label{eq:s_operator}
\end{equation}

The coherent field of the laser is taken with a mixing factor $\fmix$ that relates to the amplitude of the steady-state coherence $\braket{\sigma}$ of the QE signal. Additionally, the relative phase $\phi$ between laser and QE signal determines the interference between the two components. While in practice there is no intrinsic connection between $\sigma$ and $\beta$, we choose this relation to parametrize the amplitude and phase of the coherent state. The operator $s$ can then be used in place of $\sigma$ in any expectation value to describe the signal that is obtained if both the QE emission and coherent light of the laser is collected at the detectors. With this, we can calculate the interplay between QE signals and laser by either $G_{\text{cross},\text{co}}$ or  $G_{\text{co},\text{co}}$, as defined below. 

In general, we use low-power excitation, also known as the Heitler regime, to study the dynamics of the photon correlations under CW excitation with and without laser mixing. Further, the low excitation limit allows to better resolve the time-dependence of the multi-time correlation functions that are collected experimentally. 
\section{Revisiting the case without homodyning}

\begin{figure}
	%\centering
	\includegraphics[width=\linewidth]{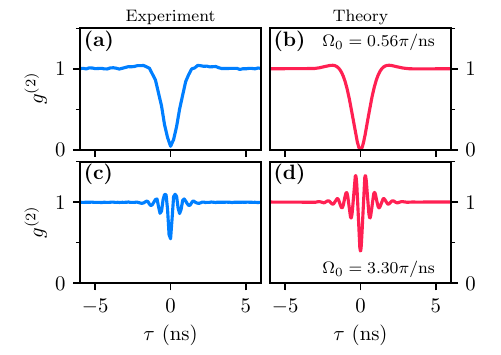}
	\caption{Second order optical coherence $g^{(2)}(\tau)$ measured between detectors $D_1, D_2$ shown in blue, theoretical values shown in red, for Rabi frequencies of (a,b) $\Omega_0=0.56\pi\,\mathrm{ns}^{-1}$ and (c,d) $\Omega_0=3.3\pi\,\mathrm{ns}^{-1}$.}
	\label{fig:g2_cw}
\end{figure}

Before focusing on the effects of mixing, it is instructive to revisit the shape of the second order correlation function $G^{(2)}$ as shown in Figure~\ref{fig:g2_cw} for two different excitation powers $\Omega_0$ of our QD. In blue we show the experimental data, as measured by the correlations between detectors $D_1$ and $D_2$, and in red the corresponding theoretical calculations. As is well known for a single emitter, we expect that $g^{(2)}(\tau=0)=0$, and the form of the curve is well-described by analytical equations \cite{Flagg2009}.

For the theoretical calculations, we include an exciton lifetime which was measured to be $T_1 = 450\,\mathrm{ps}$. In panels (a,b) we use a low power excitation of $\Omega_0=0.56\pi\,\mathrm{ns}^{-1}$, resulting in the well-known antibunching in both experiment and theory with $g^{(2)}(\tau=0)\approx 0$. 

For a higher excitation power of $\Omega_0=3.3\pi\,\mathrm{ns}^{-1}$, shown in panels (c,d), the QE undergoes Rabi-oscillations after the emission of a first photon. These are visible for short time delays in the theoretical prediction. However, the finite detector timing resolution makes the oscillations in the experimentally measured $g^{(2)}$ less pronounced and prevents $g^{(2)}(0)$ from approaching zero. In addition, dephasing of the excitonic system, often parameterized by $T_2$ of the photon, increased the damping of the oscillations. For the theory, we convolved the data with a Gaussian of \SI{250}{ps} to account for the experimental instrument response function.

%%%%%%%%%%%%%%%%%%%%%%%%%%%% Mixing g2 %%%%%%%%%%%%%%%%%%%%%%%%%%%%%%%
\section{Laser-resonance fluorescence mixing in second order coherences}
\begin{figure}
	\centering
	\includegraphics[width=\linewidth]{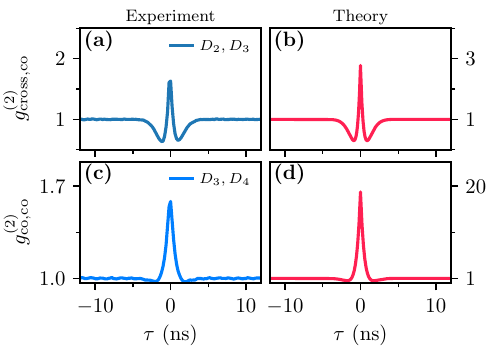}
	\caption{Second order correlation function of the resonance fluorescence and laser mixed case. In the upper panel shows the cross-co polarized case between detectors $D_1$ and $D_3$ and lower panel the co-co polarized case between $D_3$ and $D_4$ with $\Omega_0=0.3\pi\,\text{ns}^{-1}, \fmix=1,\phi=\pi$. Left: experimental data and right: theoretical calculations.}
	\label{fig:g2_crossco_coco}
\end{figure}

Now we turn our attention to the case of the $G^{(2)}$ correlation function when both laser and resonance fluorescence can be simultaneously measured on detectors. In the following we will call the signal from $D_1,D_2$ the cross-polarized signal, where no laser is mixed to the signal, while the signals at $D_3,D_4$ are co-polarized to and mixed with the laser.\\

There are two important cases: (i) the correlation between $D_1$ and $D_3$ (analog $D_2$/$D_3$), also called $g^{(2)}_{\mathrm{cross,co}}$ and (ii) the correlation between $D_3$ and $D_4$, also called $g^{(2)}_{\mathrm{co,co}}$. Examples of both cases are shown in Figure~\ref{fig:g2_crossco_coco}, with the experimental results in the left panels and the theoretical calculations on the right. 

We observe two prominent features in both cases: At $g^{(2)}(0)$ we see strong bunching, where for small $\tau$, $g^{(2)}>1$. This differs from the well-known result for resonance fluorescence $(g^{(2)}(0)=0)$ as well as that for a coherent state ($g^{(2)}(\tau) = 1$). Next, we break down the calculations to explain these observed features and then give guidance on how to control them. 
\subsection{Cross-Co Correlation}
In the case of cross-co correlations between detectors $D_1$ and $D_3$, in the $g^{(2)}(\tau)$ function (Eq.~\eqref{eq:g2}), we have to replace one $\sigma$ with the operator $s$ introduced in Eq.\eqref{eq:s_operator} yielding 
\begin{equation}
    G^{(2)}_{\mathrm{cross,co}}(t,\tau) = \braket{\sigma^{\dagger}(t)s^{\dagger}(t+\tau)s(t+\tau)\sigma(t)}.     
\end{equation}
If we break this down we obtain
\begin{align}
\begin{split}
	 G^{(2)}_{\mathrm{cross,co}}(t,\tau) 
        =&\braket{\sigma^{\dagger}(t)\sigma^{\dagger}(t+\tau)\sigma(t+\tau)\sigma(t)}\\
		&+ |\beta|^2 \braket{\sigma^{\dagger}(t)\sigma(t)}\\
		&+ 2\mathrm{Re}(\beta^*\braket{\sigma^{\dagger}(t)\sigma(t+\tau)\sigma(t)}) 
    \label{eq:parts_crossco}
\end{split}\\
        =& A+B+C \label{eq:g2parts1}
\end{align}

Here, we can identify several components: first, the original $G^{(2)}(t,\tau)$ appears in term $A$ of Eq.~\eqref{eq:parts_crossco}, which corresponds to both detectors registering a click from the resonance fluorescence. In term $B$ of Eq.~\eqref{eq:parts_crossco}, the first order correlation function at time $t$ appears, which amounts to the stationary state population, with an additional scaling by the constant factor $|\beta|^2$ depending on the amplitude of laser light. We can interpret this term as detector $D_1$ registering a photon from the resonance fluorescence, while detector $D_2$ registers a photon from the laser.

Term $C$ of Eq.~\eqref{eq:parts_crossco} contains the real part of a multi-time correlation of three $\sigma$ operators scaled with the complex factor $\beta^*$. This part can be interpreted as an interference between laser and resonance fluorescence signal. At detector $D_1$, still a QE photon is detected, while at detector $D_2$ the interference between emitter- and laser contribution to the photon state leads to a click. In the following we will see that this interference term can also take on negative values, decreasing the count rate for specific combinations of $t$ and $\tau$.

\begin{figure}
	\centering
	\includegraphics[width=\linewidth]{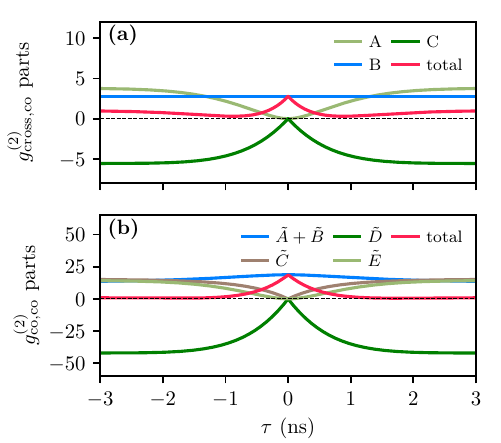}
	\caption{Different parts of (a) $g^{(2)}_{\text{cross,co}}$ as defined in Eq.~\eqref{eq:g2parts1} and (b) $g^{(2)}_{\text{co,co}}$ as defined in Eq.~\eqref{eq:g2parts} for $\Omega_0=0.3\pi\,\text{ns}^{-1}, \fmix=1,\phi=\pi$, corresponding to constant contributions as well as those of first- and second-order coherences. In panel (a), where laser photons can only be detected at one detector, the bunching at $\tau=0$ can be traced back purely to detection of laser photons at detector $D_1$. For panel (b), where laser light is mixed onto both signals, interference between dot and laser photons at both detectors can lead to additional bunching, visible by the maxima of the blue line at $\tau=0$.}
	\label{fig:g2_crossco_coco_parts}
\end{figure}

Figure~\ref{fig:g2_crossco_coco_parts}(a) shows the breakdown of the different parts according to Eq.~\eqref{eq:parts_crossco} for the case in Figure~\ref{fig:g2_crossco_coco}(b), where the individual parts are all normalized such that for the total $g^{(2)}(\tau\to\infty)=1$. 
The resonance fluorescence $g^{(2)}(\tau)$ (term $A$ in Eq.~\eqref{eq:parts_crossco}, light green line) shows the typical resonance fluorescence, low excitation behavior (cf. Figure~\ref{fig:g2_cw}(a)), reaching zero coincidences at $\tau=0$. 

In contrast, term $C$ of Eq.~\eqref{eq:parts_crossco} (dark green line) displays a completely different behavior. Still featuring zero coincidences at $\tau=0$, for the parameters chosen here, it is completely negative, interpreted as maximal destructive interference. 
Remarkably, as this part does not feature any coincidences at $\tau=0$, it can not be responsible for the bunching observed in the full $g^{(2)}$ in Figure~\ref{fig:g2_crossco_coco}.

The bunching can be explained by term B in Eq.~\eqref{eq:parts_crossco} (blue line). Intuitively, this part represents coincidences from the detection of a photon emitted by the QE at detector $D_1$, while a laser photon is detected at $D_2$. As laser and dot photons are emitted independently from each other, the process is time-independent, and this contribution appears as a line constant in $\tau$, shown in blue. Most importantly, this part is the only contribution to $g_{\mathrm{cross,co}}^{(2)}$ at $\tau=0$, meaning the bunching is made possible only by the detection of laser photons.\\
%

%%%%%%%%%%%%%%%%%%%%%%%%%%%%%%%%%%%%%%%%%%%%%%%%%%%%%%%%%%%%%%%%%%%%%%%%%%%%%%%%%%%%%%%%%%%%%%%%
\subsection{Co-Co correlation}
The picture gets even more complex if we include mixing of resonance fluorescence and laser at both detectors, $D_3$ and $D_4$. In this case, we use $s$ for all four operators, resulting in
\begin{equation}
	G^{(2)}_{\text{co,co}} = \braket{s^{\dagger}(t)s^{\dagger}(t+\tau)s(t+\tau)s(t)} \, .
\end{equation}
This multi-time operator can be expanded leading to 10 terms in total: 
\begin{equation} \label{eq:g2parts}
    \begin{aligned}
    s^{\dagger}&(t)^{\dagger}(t+\tau)s(t+\tau)s(t)=\\
    &\begin{aligned}
        	& |\beta|^{4}  \\
            +&|\beta|^2 \sigma^{\dagger}(t) \sigma(t)\\ 
	+ &|\beta|^2 \sigma^{\dagger}(t+\tau) \sigma(t+\tau)\\ 
	+ &2|\beta|^{2} Re( \overline{\beta} \sigma(t))\\ 
	+ &2|\beta|^{2} Re( \overline{\beta} \sigma(t+\tau) )\\ 
    \end{aligned}\quad&\tilde{A}\\
	&& \\
	&+2|\beta|^2Re(\sigma^{\dagger}(t+\tau) \sigma(t)) \quad&\tilde{B}\\
	&&\\
	&+  2Re(\overline{\beta}^{2} \sigma(t+\tau) \sigma(t))\quad&\tilde{C}\\ 
	&&\\
	&\begin{aligned}
    &+2Re( \overline{\beta} \sigma^{\dagger}(t) \sigma(t+\tau) \sigma(t))\\ 
	&+  2Re(\overline{\beta} \sigma^{\dagger}(t+\tau) \sigma(t+\tau) \sigma(t)) \\
	\end{aligned}\quad&\tilde{D}\\
    && \\
	&+ \sigma^{\dagger}(t) \sigma^{\dagger}(t+\tau) \sigma(t+\tau) \sigma(t)\quad &\tilde{E}
\end{aligned}
\end{equation}

As four operators ($\sigma$ or $\beta$) enter each term, in principle each of these term has the form of a $G^{(2)}$ function describing coincidences. However, single terms cannot be measured without the others and can become negative, leading to destructive interference.

The first five lines of this result, labeled $\tilde{A}$, are constant in time for CW driving, where a steady state has already been reached at $t$. Term $\tilde{B}$ resembles a $G^{(1)}$ function scaled by the amplitude of the laser $|\beta|^2$. This is the only time-dependent term that does not vanish for $\tau=0$.

Term $\tilde{C}$, instead of the usual $G^{(1)}$ function, consists of two $\sigma$ operators at different times, and vanishes at $\tau=0$.

Term $\tilde{D}$, similar to the previous case of mixing at only one detector (Term $\tilde{C}$), resembles a $G^{(2)}$ function that is missing one $\sigma$ operator, which is replaced by a $\beta$ and remains negative at all values of $\tau$. In term $\tilde{E}$, we find the usual $G^{(2)}$ function.

Again, a strong bunching around $\tau=0$ in the total correlation can be traced back to parts of the correlation function. Most of the contributions vanish around $\tau=0$, but the constant parts depending on steady-state population, coherence and laser amplitude stay finite. In contrast to the cross-co correlations, now also the time-dependent part $2|\beta|^2\mathrm{Re}[G^{(1)}(\tau)]$ appears, with finite values around $\tau=0$. Remarkably, this first-order coherence, which is a measure for the ability of light for interference \cite{gustin2018pulsed} only appears due to mixing at both detectors and corresponds to interference of laser- and dot photons at both detectors. 

Note again that the different parts in Figure~\ref{fig:g2_crossco_coco_parts} are normalized in such a way, that the total $g^{(2)}(\tau\to\infty)=1$. We stress that the normalization plays a very decisive role in the strength of the bunching. The value of the bunching for cross-co is only about $g^{(2)}_{\text{cross,co}}(0)=3$ in theory, while $g^{(2)}_{\text{co,co}}(0)$ reaches values of 20. This may be a result of phonon or charge noise in the experimental system, and will be the subject of further investigation. These values of $g^{(2)}(\tau=0)$ are greatly affected by the normalization stemming from the value of $G^{(2)}(\tau\rightarrow\infty)$. If the interference introduced by the modified $G^{(2)}$ (dark green) is maximally destructive, leading to a strong quenching of the emission for large delays, $G^{(2)}(\tau\rightarrow\infty)$ tends to very small values, in turn leading to high values of $g^{(2)}(\tau=0)$ after normalization.

\subsection{Tuning the correlation function}
\begin{figure}
	\centering
	\includegraphics[width=\linewidth]{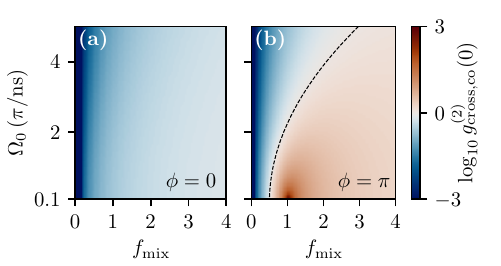}
	\caption{\textbf{Controlling the correlation function} Second order correlation function $g^{(2)}_{\mathrm{cross,co}}(\tau=0)$ on a logarithmic scale as a function of mixing strength $\fmix$ and driving strength $\Omega_0$ for a phase (a) $\phi=0$ and (b) $\phi=\pi$. Blue values correspond to anti-bunching and red values to bunching. The dashed line in (b) marks $g^{(2)}(0)=1$.}
    \label{fig:control}
\end{figure}

Having understood the origin of the signals, the next step is the control of the (anti-) bunching behavior. Besides the mixing strength between resonance fluorescence and laser $\fmix$, there are several control parameters. The driving strength $\Omega_0$ of the QE is one, that as we have seen leads to oscillatory behavior on short times. As the bunching strength is strongly dependent on the interference between the different parts in the correlation function, we tune the relative phase $\phi$ between the fluorescence and the mixed-in laser light. Experimentally, phase control can be achieved by independently mixing in some resonant laser using the HWP and QWP \cite{Schulte2015} or through wavelength detuning, as the transition imparts a varying phase shift on the scattered light which passed through zero at resonance \cite{Bennett2015}. In the experimental results presented here, the amplitude and phase of the interfering optical fields is tuned simply by adjusting linear optics in the microscope: a quarter- and half- waveplate, shown in Figure~\ref{fig1}. By changing the position of these optics, the amount of reflected laser and resonance fluorescence is altered, acting to change the relative amplitudes of the two fields. Additionally, by adjusting the incoming polarization state that interacts with the neutral exciton complex, selection rules will dictate the mixed (polarization) state that emerges in the resonance fluorescence. The orthogonal components of this mixed state experience different phase shifts when propagating back through these optics, before interfering with the reflected laser field. 

Figure~\ref{fig:control} shows the theoretical calculation of  $g^{(2)}_{\mathrm{cross,co}}(\tau=0)$ on a logarithmic scale as a function of $\fmix$ and driving strength $\Omega_0$ for (a) $\phi=0$ and (b) $\phi=\pi$. Due to the logarithmic scale, blue values mean anti-bunching ($g^{(2)}<1$) while red values correspond to bunching ($g^{(2)}>1$). Remarkably for $\phi=0$ only anti-bunching behavior is found, almost independent of the driving strength $\Omega_0$. With increasing laser mixing factor $\fmix$, the depth of the dip becomes more shallow, but stays below $1$.

For $\phi=\pi$, bunching behavior is achieved for strong enough laser mixing factor $\fmix$. The largest bunching occurs for $\fmix=1$, where the interference between the different terms is strongest. For stronger laser driving, the clear local maximum in the bunching vanishes and the regions of bunching move to larger mixing factors. This is due to the maximal destructive interference being achieved at low driving, $\fmix=1$ and $\phi=\pi$, leading to practically vanishing counts of $G^{(2)}(\tau\rightarrow\infty)$\cite{kim2024unlocking}. We conclude that by changing the mixing a control of the behavior of the correlation function is possible. 

%%%%%%%%%%%%%%%%%%%%%%%%%%%%%%%% G3 %%%%%%%%%%%%%%%%%%%%%%%%%%%%%%%%%%%%%%%%%%%%%%%%
\section{Higher order correlation function}
\begin{figure*}
	\centering
	\includegraphics[width=\linewidth]{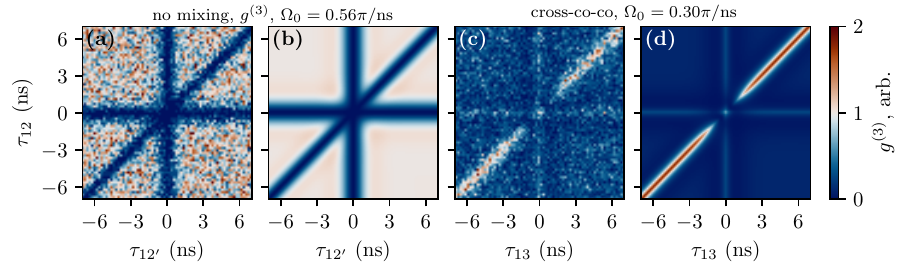}
	\caption{(a,b) measured and calculated $g^{(3)}$ without laser mixing, i.e., $\fmix=0$, for a Rabi frequency of $\Omega_0=0.56\pi\,\mathrm{ns}^{-1}$. The prominent feature is antibunching along the $\tau$ axes as well as the diagonal. (c,d) mixing of QD signal with laser for $\fmix=1$ and $\phi=\pi$ on detectors $D_3$ and $D_4$. At detector $D_1$, still only the QD signal is measured. In contrast to (a,b), strong bunching along the diagonal as well as moderate bunching along the $\tau$ axes, especially at the center, is now visible.}
	\label{fig:g3_cw}
\end{figure*}

We can also go beyond the second order correlation function and experimentally detect the third order correlation function. The results of this are shown in Figure~\ref{fig:g3_cw} as maps of the measured and calculated $g^{(3)}$-functions, depending on the arrival times of photons at two detectors, after a photon has been registered at detector $D_1$. In panels (a,b), the we use a Rabi frequency of $\Omega_0=0.56\pi\,\mathrm{ns}^{-1}$ and three detectors which sample the cross-polarized photons, which we call $D_1$, $D_2$ and $D_{2'}$. A cross-shaped anti-bunching can be seen for $\tau_{12}=0$ and $\tau_{12'}=0$, which means that no coincidences are detected for zero time delay between the first photon detection at $D_1$ and either $D_2$ or $D_{2'}$. Additionally, along the diagonal of the plot anti-bunching is visible corresponding to a zero time delay between events on detectors $D_2$ and $D_{2'}$. The same features visible in the experimental data can also be seen in numerical calculations shown in panel (b).

Next, in panel (c,d) the results for $g^{(3)}$ under mixing of the QE signal with the laser is shown. Here, we choose $\fmix=1$ and $\phi=\pi$, which when looking at Eq.~\eqref{eq:s_operator} leads to maximal destructive interference between laser and the resonance fluorescence component. It is immediately visible that the shape of the acquired $g^{(3)}$ function changes when compared to panels (a,b), with bunching clearly visible along the diagonal as well as the $\tau$ axes, especially at $\tau_{12}=\tau_{13}=0$. Note that the plots in panel (c,d) are not normalized.

It is noteworthy that we can further define the $n$th order optical coherence including mixing at all detectors as $G^{(n)}(t,\tau=0) = \braket{(s^{\dagger})^n(t) s^n(t)}$. Expanding this expression leads to
\begin{align}
	\begin{split}
			G^{(n)}(t,\tau=0) = &|\beta|^{2n} + n^2|\beta^{2(n-1)}|\braket{\sigma^{\dagger}(t)\sigma(t)} \\
		&+ 2n|\beta|^{2n-1} |\!\braket{\sigma(t)}\!| \cos \phi.
	\end{split}
\end{align}

Again, all expectation values containing more than one $\sigma$ operator at the same time vanish, leaving us with components of the coherent laser state mixed with the resonance fluorescence signal coming from the QE, revealing that for all orders of correlation functions, laser photons are necessary for detecting events at $\tau=0$. 

%%%%%%%%%%%%%%%%%%%%%%%%%%%%%%%% Conclusion %%%%%%%%%%%%%%%%%%%%%%%%%%%%%%%%%%%%%%%%%%%%%%%%
\section{Conclusions}
In conclusion, we have measured and theoretically analyzed multi-photon correlation functions for the case of resonance fluorescence being mixed with a coherent laser state. Segmenting the correlation function into several parts, we traced back the origin on bunching to an interplay of interference between laser and QD photons and normalization of the signals to long time steady states. Our theoretical investigations clarify the origin of the measured signals, keeping not only the resonance fluorescence interpretation intact but furthermore allowing us to show the tunability of the multi-photon correlation via adjustment of the relative phase, driving strength, and mixing factor.

Beyond the second-order correlation function, we also measured and calculated the third-order correlation function, finding again a mixture of anti-bunching and bunching behavior when the signal is mixed with a laser pulse. Extension of this analysis to higher-order correlations is possible but at a cost of increased integration time for experimental data.

The experimental results presented in this work demonstrate the capability to produce tunable photon statistics by interfering optical fields of varying amplitude and phase using linear optics. The agreement between theoretical model and experimental results provides a link between the measurements, and an understanding of how the interfering optical fields contribute to the resulting correlation functions. Such a technique offers insights into fundamental quantum optics surrounding resonance fluorescence and photon correlation, and an experimental testbed for non-Gaussian state engineering.

\section{Acknowledgments}

Cardiff University acknowledges financial support from EPSRC Grant No. EP/T001062/1, EP/Z53318X/1 and EP/T017813/1. T.K.B. and M.C. are supported by the Return Program of the State of North Rhine-Westphalia. The experimental datasets used and/or analysed during the current study are available from the Cardiff University Data Repository DOI: 10.17035/cardiff.30245599. 

\newpage
\bibliography{main_arxiv}
\end{document}